\documentclass[%
 reprint,
 amsmath,amssymb,
 aps,
]{revtex4-2}

\usepackage{graphicx}
\usepackage{dcolumn}
\usepackage{bm}

\def\diff{\textrm{d}}

\def\tz{{\tt z}}
\def\dj{d\kern-0.4em\char"16\kern-0.1em}
\def\Dj{\mbox{\raise0.3ex\hbox{-}\kern-0.4em D}}

\begin{document}

\title{The role of torsion in holographic conductivity}

\author{Du\v{s}an \Dj or\dj evi\'{c}}
\email{dusan.djordjevic@ff.bg.ac.rs}
\author{Ivana \Dj or\dj evi\'{c}}%
\author{Aleksandra Go\v{c}anin}
\author{Dragoljub Go\v{c}anin}
\email{dragoljub.gocanin@ff.bg.ac.rs}
\affiliation{%
Faculty of Physics, University of Belgrade\\ Studentski Trg 12-16, 11000 Belgrade, Serbia
}%





\begin{abstract}
Generalizing the usual setup for holographic duality, where bulk spacetime is described by pseudo-Riemannian geometry, we consider a Riemann-Cartan bulk with non-trivial torsion as a background for an electromagnetic gauge field dual to $U(1)$ boundary current. Working in the probe limit, we explore how the bulk torsion, which induces spin current at the boundary, affects the electric conductivity of the boundary theory. We consider standard types of non-minimal couplings between torsion and the electromagnetic field found in the literature, and the results indicate that these torsion couplings are more suitable candidates, compared to the common minimal coupling regime, for a holographic description of the existing experimental findings regarding conductivity.
\end{abstract}

\maketitle


\section{Introduction \label{introduction}}

General relativity (GR) describes spacetime as a pseudo-Riemannian manifold whose affine structure, including connection and curvature, is completely determined by the metric tensor. Thus, the torsion tensor vanishes, and the dynamics of the gravitational field, as well as the dynamics of matter fields, is dictated solely by the metric. Although it is notoriously challenging to promote GR into quantum theory, one feature quantum gravity is generally believed to exhibit is that it should be holographic, meaning that its degrees of freedom pertain to its boundary. This is most explicitly stated in the AdS/CFT correspondence, which emerged from string theory \cite{Maldacena:1997re}, and was later generalized to gauge/gravity duality. The assumption behind this duality is that there is an equivalent description of a (quantum) gravity theory in asymptotically anti-de Sitter (AdS) spacetime in terms of a quantum field theory (QFT) associated with the asymptotic boundary. \\

Although one commonly assumes the GR framework, not all theories of gravity demand that spacetime has vanishing torsion. For example, in order to formulate gravity as a gauge theory, one has to consider connection to be independent from the metric. A particular example is the Poincare gauge theory (PGT) \cite{blagojevic2001gravitation}, where gravity is introduced by localizing the Poincare symmetry. As connection is not \textit{a priori} determined from the metric, torsion is generically part of the system, and spacetime is described by Riemann-Cartan geometry. Despite its name, PGT has certain differences from the standard Yang-Mills theory and fails short of being a genuine gauge theory of gravity based on a fiber bundle structure (except in three dimensions). There is, however, a way to formulate gravity as a genuine gauge theory in any odd number of dimensions, by describing it in terms of Chern-Simons (CS) forms \cite{Chamseddine:1989nu, hassaine2016chern}. In three dimensions, CS gravity without matter fields is classically equivalent to GR with cosmological constant. However, five-dimensional CS gravity differs from its GR counterpart. Compared to the GR action, the CS action in five dimensions contains an additional term quadratic in curvature, and it amounts to a very special case of the Lovelock action with an enhanced (A)dS (depending on the sign of the cosmological constant) gauge symmetry. Moreover, the equations of motion of five-dimensional CS gravity do not imply zero torsion, even in the matter-free case. 

Holography for a Riemann-Cartan spacetime was considered, to the best of our knowledge, for the first time in \cite{Banados:2006fe}, using five-dimensional CS gravity as an example. Subsequent work \cite{Klemm:2007yu} analyzed the case of three-dimensional gravity with torsion, again based on a CS description. Another approach to Riemann-Cartan holography was introduced in \cite{Leigh:2008tt, Petkou:2010ve} for a four-dimensional bulk. Subsequent work on three-dimensional gravity includes \cite{Blagojevic:2013bu}, while a generalization of CS gravity to any odd number of dimensions was done in \cite{Cvetkovic:2017fxa}. In \cite{Gallegos:2020otk}, one finds an important generalization of the previous discussion to holographic hydrodynamics with a spin current, together with some insights regarding boundary terms. Further discussion about boundary terms in the context of AdS/CFT duality with torsion can be found in \cite{Erdmenger:2022nhz}, while \cite{Dordevic:2024ziw} discussed AdS/BCFT duality with a Riemann-Cartan bulk. Torsion also plays an important role in the holographic description of Weyl semi-metals with dislocations \cite{Juricic:2024tbe}. It is worth noting that this list is quite small compared to the rest of the literature involving gauge/gravity duality, where, standardly, one postulates torsion-free pseudo-Riemannian geometry. With this paper, we open a new direction of research regarding holography with torsion and its condensed matter applications, focusing particularly on holographic conductivity. 

The paper is organized as follows. In section \ref{Cs} we review five-dimensional CS gravity and comment on its holographic properties. We focus on a particular black hole solution with torsion and compute the one-point function of the spin current in the dual QFT. In Section \ref{U1} we add a $U(1)$ gauge field in the bulk and consider its couplings with torsion that lead to holographic conductivity. We summarize our results and offer some perspective in Section \ref{D&C}. 

\section{Chern-Simons gravity}\label{Cs}
Five-dimensional AdS CS gravity is defined by the following $SO(4,2)$ gauge invariant action, 
\begin{align}
   &S_{\textrm{CS}}^{(5)}= \frac{k}{8}\int\varepsilon_{ABCD}\bigg(\frac{1}{l}R^{AB}\wedge R^{CD}\wedge e^{E}\\
   &+\frac{2}{3l^{3}}R^{AB}\wedge e^{C}\wedge e^{D}\wedge e^{E}+\frac{1}{5l^{5}}e^{A}\wedge e^{B}\wedge e^{C}\wedge e^{D}\wedge e^{E}\bigg),\nonumber
\end{align}
where $k$ is a dimensionless constant (the CS level) and $l$ is the AdS radius. There are two dynamical fields, the vielbein $e^{A}$ and the spin-connection $\omega^{AB}$, while $R^{AB}=\diff\omega^{AB}+\omega^{A}_{\;\;C}\wedge\omega^{CB}$ is the curvature.
Equations of motion imply 
that the torsion $2$-form $T^{A}=\mathrm{d}e^{A}+\omega^{A}_{\;\;B}\wedge e^{B}$ does not necessarily vanish. For example, one particular black hole solution with non-vanishing torsion is given by the metric 
\begin{equation}\label{metrika}
    \diff s^2=-\left( \frac{r^2}{l^2}-\mu \right)\diff t^2+\frac{\diff r^2}{\left( \frac{r^2}{l^2}-\mu \right)}+r^2\diff \Sigma_3^2,
\end{equation}
and the torsion
\begin{align}\label{torzija}
T^{0}&=T^{1}=0,\nonumber\\
    T^i&=-\frac{\delta}{r}\varepsilon_{ijk}e^j\wedge e^k,
\end{align}
with $\delta$ arbitrary constant that measures the strength of torsion \cite{Canfora:2007xs}. Indices $i,j,k$ take values from $\{2,3,4\}$, and we will not distinguish between lower and upper Lorentz $i,j,k$ 
indices. We set $l=1$. As torsion is not identically equal to zero on-shell, spin-connection and vielbein can indeed be considered as independent fields. Interestingly, there are no restrictions for the three-manifold $\Sigma_{3}$, apart from the fact that it is fixed for all values of the radial coordinate. However, presumably, the most interesting case for applications is that of a planar black holes, where $\diff \Sigma_3^2=\diff x^2+\diff y^2 +\diff z^2$ and (\ref{metrika}) induces flat boundary metric. Having in mind applications to condensed matter systems, namely, the holographic conductivity, this is the case we will consider.  

Now we turn to holography. Using symmetries and equations of motion, one can choose the appropriate Fefferman-Graham (FG) gauge for the bulk vielbein and spin-connection as follows,
\begin{align}\label{FGexp}\nonumber
e^1&=-\frac{\diff \rho}{2\rho}, \hspace{0.5cm}e^a=\frac{1}{\sqrt{\rho}}(\overline{e}^a+\rho \overline{k}^a),\\
\omega^{a1}&=\frac{1}{\sqrt{\rho}}(\overline{e}^a-\rho \overline{k}^a),\hspace{0.4cm}
\omega^{ab}=\overline{\omega}^{ab},
\end{align}
where boundary fields are denoted by a bar sign and ``non-radial" indices $a$, $b$ take values from $\{0,2,3,4\}$; index 1 is the radial index. 
Note that the FG expansion of the bulk fields in powers of the FG radial coordinate $\rho$ is finite, which is a peculiarity of CS gravity. 

The holographic prescription dictates the following (schematic) relation,
\begin{equation}
    e^{iW}=\int \mathcal{D}e \mathcal{D}\omega\; e^{iS}\approx e^{i S_{\text{on-shell}}}.
\end{equation}
The second (approximate) equality follows from the fact that we treat gravity classically (saddle-point approximation). The correlation functions, normally generated by the functional $W$ of the boundary CFT, are holographically obtained from the on-shell gravity action. However, as the AdS spacetime is noncompact, infinities arise when one integrates over the radial coordinate up to the conformal boundary $\rho=0$, and therefore an appropriate holographic renormalization has to be performed. This resembles the well-known renormalization procedure that has to be performed in QFT to obtain finite correlation functions. After removing infinities and defining the renormalized action $S_{ren}$, one has 
\begin{equation}\label{del}
    \delta S_{ren}=\delta W= \int_{\partial \mathcal{M}}\left(\delta \overline{e}^a\wedge \tau_a+\frac{1}{2}\delta\overline{\omega}^{ab}\wedge s_{ab}\right).
\end{equation}
Generally, (\ref{del}) does not hold without additional finite boundary terms, and the importance of these terms was emphasized in \cite{Gallegos:2020otk}.
The holographic dictionary interprets $\tau_a$ as $\langle \mathcal{T}_a\rangle_{QFT}$, the expectation value of the stress-energy tensor of the boundary QFT, and $s_{ab}$, according to \cite{Banados:2006fe,Cvetkovic:2017fxa}, as the expectation value of the spin current $\langle\mathcal{S}_{ab}\rangle_{\mathrm{QFT}}$. The existence of spin current in the boundary theory is a novelty of gravity with torsion, and its importance was realized in \cite{Gallegos:2020otk}. 
For five-dimensional CS gravity, general formula for the one-point function of a spin current is given by \cite{Banados:2006fe}
\begin{equation}   \langle\mathcal{S}_{ab}\rangle_{\mathrm{QFT}}=2k\varepsilon_{abcd}\overline{T}^c\wedge \overline{k}^d.
\end{equation}
The spin current is proportional to the boundary torsion, and therefore its existence is conditioned on having bulk torsion, at least in the holographic setting. Note that torsion has received attention within the condensed matter community, where it usually plays the role of dislocations \cite{deJuan:2009ldt, Krner1994CrystalLD, Huang2019NiehYanAT}.  The discussion of spin current also includes \cite{Cartwright:2024dcj}. 

We can now compute the one-point correlation function of a spin current at the boundary for the previously introduced black hole with torsion. The result is (see Appendix \ref{AppA} for details) 
\begin{align}
\langle\mathcal{S}_{0i}\rangle_{\mathrm{QFT}}&=0, \\ 
\langle\mathcal{S}_{ij}\rangle_{\mathrm{QFT}}&=k\mu\delta \;\diff x^i\wedge \diff x^j\wedge \diff t.
\end{align}
Note that the spin current is proportional to the temperature squared, since $\mu\sim T^{2}$. In condensed matter physics, especially spintronics, spin current plays an important role in spin transport phenomena. However, the definition of a spin current used here is a relativistic one, and one should seek to understand more about the nature of this spin current in order to relate it to the one traditionally used in spintronics, see \cite{an2011universaldefinitionspincurrent}.

\section{Adding electromagnetic field}\label{U1}

So far, our discussion has been limited to pure CS gravity. A more realistic version of this theory should also include matter fields. 
Let $A=A_\mu \diff x^\mu$ be a gauge field in the bulk for the $U(1)$ gauge group, and $F=\diff A$ the associated field strength. It is obvious that the standard minimal coupling action $-\frac{1}{2e^{2}}\int F\wedge \star F$ does not couple the gauge field $A$ with torsion. The action only involves the metric, and it is therefore irrelevant whether our spacetime is pseudo-Riemannian, with a Levi-Civita connection and no torsion, or Riemann-Cartan, with non-trivial torsion. From a holographic point of view, this means that the boundary electromagnetic current is completely blind to the previously discussed spin current (at least in the probe limit, see below) which is unlikely to be relevant in the real world given the interpretation of the boundary torsion as dislocations in the material. Moreover, holographic systems in the minimal coupling regime similar to the one at hand have been studied in the past, and it is known that, typically, they do not capture the characteristic Drude peak in the optical conductivity. With this in mind, we can proceed to build a holographic system that has a coupling between the $U(1)$ gauge field and torsion in the bulk. 

To address the question of possible couplings between the gauge field $A$ and torsion, we start with the standard action $-\frac{1}{2e^{2}}\int F\wedge \star F$ and make the following substitution, 
$\label{kap1}
    \star F \rightarrow (1-\lambda^{2}\star(T^A \wedge \star T_A))\star F$, where $\lambda^{2}$ is a positive coupling constant. Similar couplings, but in four dimensions, were studied in \cite{Rubilar:2003uf}.
The bulk physics is thus described by the action
\begin{equation}\label{CS_EM_T}
  S_{\textrm{CS}}^{(5)}  -\frac{1}{2e^2}\int F\wedge (1-\lambda^{2}\star(T^A \wedge \star T_A))\star F.
\end{equation}
We will be working in the probe limit where one can neglect the backreaction of the gauge field on geometry. It is worth noting that we are not aware of any full backreacting black hole solution of a five-dimensional CS gravity coupled to electromagnetism that involves torsion, even for $\lambda^{2}=0$.  
However, as the full backreacting theory is expectedly hard, and holographic renormalization of that (torsionful) theory has not been conducted in the literature (the lack of symmetry would prevent us from using results of \cite{Banados:2006fe}), we will assume that it makes sense to work in the standard probe limit, having in mind that this limit may have to be realized in some theory similar to, yet different from CS gravity. 

Equations of motion for the gauge field are given by 
\begin{equation}
    \partial_{\mu}\left(\sqrt{-g}\mathcal{C}F^{\mu\nu}  \right)=0,
\end{equation}
where $\mathcal{C}=1-\lambda^{2}\star(T^A \wedge \star T_A)=1+\frac{3\lambda^{2}\delta^2}{r^2}$ for the black hole (\ref{metrika})-(\ref{torzija}). We choose a gauge $A_r=0$. Decomposing the free index $\nu$ into $(r,\alpha)$, we respectively obtain the following system of equations: 
\begin{align}\label{prvaek}
    \partial_{\alpha}F^{\alpha r}&=0,\\
    \partial_r \left(\sqrt{-g}\mathcal{C}\frac{(r^2-\mu)}{r^2}\partial_r A_\alpha\right)&+\sqrt{-g}\mathcal{C} \partial_{\beta}F^{\beta \alpha}=0.
\end{align}
From the second equation we find that, asymptotically, $A_{\alpha}$ behave as 
\begin{equation}\label{Aasym}
    A_{\alpha}\sim A_\alpha^{(0)} + B_{\alpha}\frac{\ln r}{r^2} + \frac{A_{\alpha}^{(1)}}{r^2},
\end{equation}
where $B_{\alpha}$ depends only on $A_\alpha^{(0)}$. The field $A_\alpha^{(0)}$ is kept fixed at the boundary and corresponds to the source of the boundary $U(1)$ current. Generally, boundary fields $A_\alpha^{(0)}$ and $A_{\alpha}^{(1)}$ can depend on all of the $t,x,y,z$ coordinates, though here we will consider only symmetric configurations with no explicit dependence on the boundary spatial directions. We will see below that the one-point function of the boundary current is given by $A_{\alpha}^{(1)}$, and therefore (\ref{prvaek}) implies that the $U(1)$ current is conserved.

The coupling between $U(1)$ gauge field and torsion is more restrictive than its coupling with the curvature, as torsion is a tensor of rank three, and it is harder to construct a gauge invariant Lagrangian that does not involve higher powers of torsion. Apart from the coupling already considered, we can also take the so called Preuss coupling \cite{Rubilar:2003uf} given by
\begin{equation}\label{alter} 
    \int \left(T^A\wedge F  \right)\wedge \star\left(T_A\wedge F  \right).
\end{equation}
We will comment on this alternative choice below. Note that the most general Lagrangian is presented in \cite{Itin:2003hr}. 

\subsection{Holographic conductivity}
Let us first discuss the $\lambda^{2}=0$ case, where there is no coupling between the torsion and the $U(1)$ gauge field in the bulk. We want to turn on the electric field in the $x$ direction at the boundary. To do so, we solve for the perturbation $\delta A=e^{-i\omega t}A_x(r)\diff x$ and obtain the following differential equation,
\begin{equation}
    \frac{\diff^2 A_x}{\diff r^2}+ \frac{3 r^2- \mu}{r \left(r^2-\mu \right)}\frac{\diff A_x}{\diff r}+\frac{\omega^2}{(r^2-\mu)^2}A_x=0.
\end{equation}
This equation can be solved analytically and the solution that assumes the in-falling boundary conditions at the horizon is given by a particular hypergeometric function, see (\ref{Hiep}) in Appendix \ref{AppB}. 

Using linear response theory, we can obtain the conductivity in the boundary QFT by computing the ratio of the current one-point function and the applied background electric field, or equivalently, by using the current-current two-point function. We choose the former logic, and obtain 
\begin{equation}\label{holcond}
    \sigma(\omega)\equiv\sigma_{xx}(\omega)=\frac{2}{i\omega}\frac{A_{x}^{(1)}}{A_{x}^{(0)}}+i\frac{\omega}{2}.
\end{equation}
Note that, since our coupling between the $U(1)$ gauge field and torsion is of the dilaton type, analogously as for the class of Einstein-Maxwell-Dilaton theories (compare with \cite{Dehyadegari:2017fqo, Liu:2022bdu}), the formula (\ref{holcond}) for holographic conductivity can be derived by following the standard prescription of holographic renormalization and applying Ohm's law on the current one-point function and the background electric field at the boundary.   

By expanding the solution (\ref{Hiep}) near the boundary, we can obtain an analytic expression for the holographic conductivity. The result is 
\begin{align}\label{condntor}
    \sigma(\omega)= i \omega\left ( \psi\left(1-\frac{i \omega }{2
   \sqrt{\mu }}\right)+\gamma\right)  +\sqrt{\mu},
\end{align}
where $\psi(x)=\frac{\diff \ln \Gamma(x)}{\diff x}$ is the digamma function, and $\gamma\approx 0.577216$ is Euler's constant. We see that the DC conductivity is finite and has the expected linear temperature dependence, $\sigma_{DC}=\sigma(0)=\sqrt{\mu}=2\pi T$. Quazinormal modes for this type of black hole were considered in \cite{Gomez-Navarro:2017fyx}, so it is not surprising that an analytic expression for conductivity can be found. Note, however, that, even though generally $\sigma(\omega)$ can be found only numerically, there are cases where analytic form of the holographic conductivity is found, see, for instance, \cite{Ren:2021rhx}.\\

\begin{figure}[h!] 
    \centering
\includegraphics[width=\linewidth]{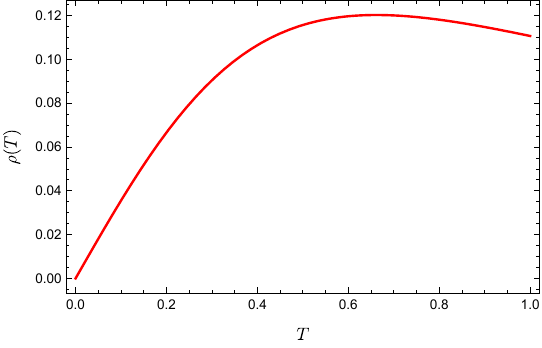}
    \caption{DC resistivity versus temperature, for fixed $\delta$ and $\lambda$.}
    \label{Fig1}
\end{figure}

\begin{figure}[h!]
    \centering
\includegraphics[width=\linewidth]{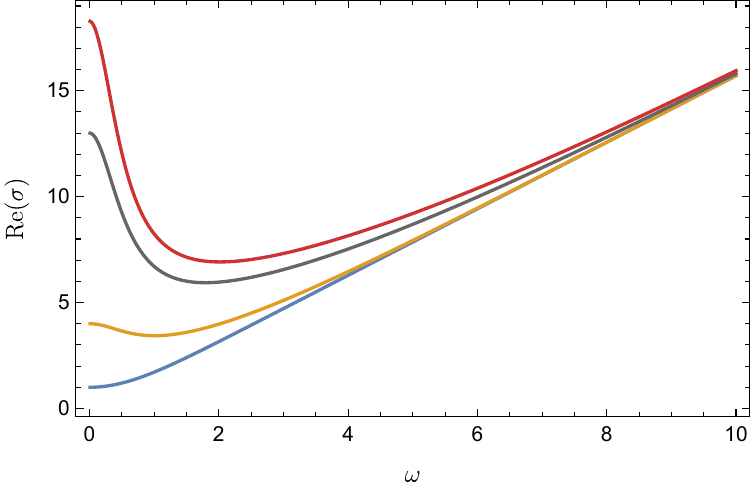}
    \caption{Real part of the conductivity $\sigma$. In the numerical calculation, temperature is set to $T=2\pi$. Lines in the bottom-up order correspond to $\lambda\delta=0$ (blue), $\lambda\delta=1$ (orange), $\lambda\delta=2$ (grey), $\lambda\delta=2.4$ (red), respectively. Color online. }
    \label{Fig2}
\end{figure}

\begin{figure}[h!] 
    \centering
\includegraphics[width=\linewidth]{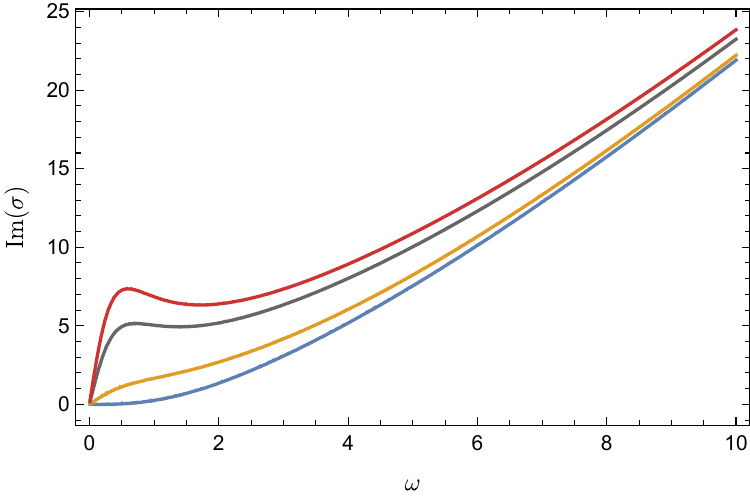}
    \caption{Imaginary part of the conductivity $\sigma$. In the numerical calculation temperature is set to $T=2\pi$. Lines in the bottom-up order correspond to $\lambda\delta=0$ (blue), $\lambda\delta=1$ (orange), $\lambda\delta=2$ (grey), $\lambda\delta=2.4$ (red), respectively. Color online. }
   \label{Fig3}
\end{figure}

Of course, we are more interested in the effects of torsion. As we have pointed out, for torsion to be relevant for the boundary theory, we have to consider a non-minimal coupling ($\lambda^{2}\neq 0$) between the gauge field and torsion in the bulk. In the case of (\ref{CS_EM_T}) we again come to a (more complicated) differential equation for $A_{x}(r)$, see (\ref{nmdiff}) in Appendix \ref{AppC}, which, expectedly, does not have an analytical solution and demands numerical tools. 
Note also that now there are two important scales in the problem. One is set by the black hole temperature $T=\frac{\sqrt{\mu}}{2\pi}$ and corresponds to the temperature of the dual QFT, and the other one, $\delta$, is set by the spin current expectation value, or equivalently, by the strength of the torsion. The DC conductivity is now 
\begin{equation}\label{DCtor}
\sigma_{DC}=\sqrt{\mu}+ \frac{3\lambda^{2}\delta^2}{\sqrt{\mu}}.
\end{equation}
We can define resistivity in terms of temperature $T=\frac{\sqrt{\mu}}{2\pi}$ as $\rho(T)=1/\sigma_{DC}$, and plot it, see Figure \ref{Fig1}. Note that the plot starts at zero value of resistivity at zero temperature and increases up to certain temperature. This is a characteristic behavior for metals. However, at $T_c=\frac{\sqrt{3}\lambda \delta}{2\pi}$, the resistivity starts to decrease with increasing temperature, indicating semiconductor behavior. 
As for the AC conductivity, we can compute it using numerical techniques. We apply the infalling boundary conditions at the horizon and use the NDSolve function from Mathematica to numerically solve the differential equation (\ref{nmdiff}) and read off the asymptotic behavior of the field $A_x$ near the boundary, see Figures \ref{Fig2} and \ref{Fig3}. 

We can immediately note some characteristics of the plotted graphs. First, for $\omega=0$ we confirm the analytic result for DC conductivity (\ref{DCtor}). Furthermore, the case $\delta=0$ corresponds to the analytic formula (\ref{condntor}) that we have previously obtained. Indeed, we can see from the graph the $\sim\omega^2$ behavior near $\omega=0$, and $\sim \omega$ behavior for large frequencies. The latter is a consequence of conformal invariance in five dimensions and also survives for $\delta\neq 0$ when there are dislocations in the boundary systems sourcing the spin current. However, for $\delta\neq 0$ the low-frequency regime drastically changes. As discussed in the introduction, the $\delta=0$ case does not capture the expected Drude peak, which should be present in this regime. However, we see that nonzero $\delta$ (torsion in the bulk) gives us precisely this peak, with the peak being greater if the parameter $\lambda\delta$ is increased. Let us also point out that the form of the real part of conductivity, especially for $\lambda\delta\approx 2.5$, matches the experimental results discussed in \cite{xu2022linear} where the conductivity properties of $\mathrm{Ir}_2\mathrm{In}_8\mathrm{Se}$ were experimentally tested. Actually, one can see the similarity between the $\rho(T)$ dependence experimentally obtained in \cite{xu2022linear} and the one we derived from our holographic model. This similarity with the experimental data is valid only below the critical temperature $T_c$ (apart from small corrections for very low temperatures). However, this transition between metallic and semiconductor electrical conductivity might signal that the boundary theory indeed describes a semimetal, see for instance the conductivity versus temperature behavior of certain semimetals in \cite{Chen2015OpticalSS, Pavlosiuk2015AntiferromagnetismAS}. Finally, let us just point out that the alternative Preuss coupling (\ref{alter}) gives us essentially the same results for holographic conductivity, supporting the robustness of the effect.  

\section{Discussion and conclusion}\label{D&C}
Holographic duality for Riemann-Cartan spacetime in the probe limit can be formulated, and the results indicate that a non-minimal coupling between $U(1)$ gauge field and spacetime torsion is more promising than the usual minimal coupling. An important issue that has not been addressed here is backreaction. Furthermore, holographic renormalization, together with the calculation of the spin current two-point function, has yet to be performed. An investigation that could follow this work is the study of holographic superconductivity and the influence of torsion on the critical temperature.  

In our model, the DC conductivity is finite, and its behavior is drastically different when $\lambda^{2}\neq 0$. As for the AC conductivity,
the Drude peak is present in real materials, and holography should be able to capture it \cite{Horowitz:2012ky}. We have demonstrated that the proposed model with non-minimal coupling between the electromagnetic gauge field and torsion conforms to the existing experimental findings. Still, we are not claiming this to be the only possible explanation. For example, one could also consider a non-minimal coupling between the gauge field and the Riemannian curvature, mediated by the Ricci scalar, which in our case is given by $R=-20+\frac{6 \mu }{r^2}$, and therefore can reproduce a similar plot for conductivity. However, in the case of curvature couplings and zero torsion, there is only one relevant scale - the one set by the temperature, and we cannot, with fixed parameters in the bulk action, tune the intensity of the Drude peak. Moreover, for the DC conductivity, the torsion parameter $\delta$ allows us to tune the semimetal behavior. Finally, let us mention that the boundary dual of five-dimensional CS gravity is a non-unitary theory \cite{Banados:2004zt}, and it would be interesting to understand the consequences of this fact on the model considered in this paper.

\begin{acknowledgments}
We thank Mihailo Cubrovic for sharing his knowledge and Mathematica codes for numerical computation of the optical conductivity.
Work of D. \Dj., I. \Dj., A. G. and D. G. is supported by the funding provided by the Faculty of Physics, University of Belgrade, through grant number 451-03-47/2023-01/200162 by the Ministry of Science, Technological Development and Innovations of the Republic of Serbia. The research was supported by the Science Fund of the Republic of Serbia, grant number TF C1389-YF, Towards a Holographic Description of Noncommutative Spacetime: Insights from Chern-Simons Gravity, Black Holes and Quantum Information Theory - HINT.
\end{acknowledgments}

\bibliographystyle{elsarticle-num} 
\bibliography{ref}

\appendix

\section{Spin current}\label{AppA}

We can compute the one-point function of the spin current at the boundary for the black hole with torsion (\ref{metrika})-(\ref{torzija}) by changing the radial coordinate $r$ to the FG coordinate $\rho$, defined as \cite{Dordevic:2023crm, Juricic:2024tbe}
\begin{equation}
r=\frac{1}{\sqrt{\rho}}+\frac{\mu}{4}\sqrt{\rho},
\end{equation}
and rewrite the black hole metric (\ref{metrika}) in the FG form 
\begin{equation}\label{metricFG}
    \diff s^2=\frac{\diff \rho^2}{4\rho^2}+\frac{\eta_{ab}}{\rho}\Big(\overline{e}^a_\alpha \overline{e}^{b}_{\beta}+\rho (\overline{e}^a_\alpha \overline{k}^{b}_{\beta}+\overline{e}^a_\beta \overline{k}^{b}_{\alpha})+\rho^2\overline{k}^a_\alpha \overline{k}^{b}_{\beta}\Big)\diff x^\alpha \diff x^\beta.
\end{equation}
Now we can directly read off all the boundary fields appearing in the FG expansion (\ref{FGexp}), and compute the promised one-point function. For the black hole at hand, we have the vielbein $\overline{e}^{a}=\diff x^\mu \delta_\mu ^a$, nonzero components of the spin-connection are  $\overline{\omega}^{ij}=\varepsilon_{ijk}\overline{e}^k$,
while $\overline{k}^a=\epsilon\frac{\mu}{4} \delta_\mu^a\diff x^\mu$, where $\epsilon=-1$ if $a=0$ and $\epsilon=1$ otherwise. It is now easy to compute components of the boundary torsion, 
\begin{equation}
\overline{T}^0=\diff\overline{e}^0+\overline{\omega}^{0}_{\;\;i}\wedge\overline{e}^i=0, 
\end{equation}
while for the spatial components we have
\begin{equation}
    \overline{T}^i=\delta \varepsilon_{ijk}\diff x^j\wedge \diff x^k.
\end{equation}
This immediately implies that $\langle\mathcal{S}_{0i}\rangle_{\mathrm{QFT}}=0$, while nonzero components are given by 
\begin{equation}
    \langle\mathcal{S}_{ij}\rangle_{\mathrm{QFT}}=k\mu\delta \;\diff x^i\wedge \diff x^j\wedge \diff t.
\end{equation}
Taking the Hodge dual of the last expression we have 
\begin{equation}
    \star  \langle\mathcal{S}_{ij}\rangle_{\mathrm{QFT}}=k\mu\delta\varepsilon_{ijk}\diff x^k.
\end{equation}
From the spin current we can also compute the axial current, as advertised in \cite{Banados:2006fe,Juricic:2024tbe}, to obtain  
\begin{equation}
\mathcal{J}_{\mathrm{ax}}^0=k\mu\delta, \hspace{9mm}\mathcal{J}_{\mathrm{ax}}^i=0.
\end{equation}

\section{Minimal coupling conductivity}\label{AppB}

This is a simpler case where there is no coupling between the gauge field and the torsion ($\lambda^{2}=0$). We turn on the electric field in the, for example, $x$ direction at the boundary. To do so, we solve the equations of motion for the perturbation $\delta A=e^{-i\omega t}A_x(r)\diff x$ and obtain the following differential equation,
\begin{equation}
\frac{\diff^2 A_x}{\diff r^2}+ \frac{3 r^2- \mu}{r \left(r^2-\mu \right)}\frac{\diff A_x}{\diff r}+\frac{\omega^2}{(r^2-\mu)^2}A_x=0.
\end{equation}
This equation can be solved analytically. Making the substitution $r=u\sqrt{\mu}$, we have
\begin{equation}
    \frac{\diff^2 A_x}{\diff u^2}+ \frac{3 u^2- 1}{u \left(u^2-1 \right)}\frac{\diff A_x}{\diff u}+\frac{\omega^2}{\mu(u^2-1)^2}A_x=0,
\end{equation}
and the solution is given by 
\begin{align}\label{Hiep}
   A_{x}(u)&=(u-1)^{\frac{-i \omega }{2\sqrt{\mu}}} (u+1)^{\frac{-i \omega }{2\sqrt{\mu}}}  \nonumber\\
   &\times \, _2F_1\left(1-\frac{i \omega}{2\sqrt{\mu}} ,\frac{-i \omega }{2\sqrt{\mu}};\frac{-i
   \omega}{\sqrt{\mu}} +1;1-u^2\right),
\end{align}
where $_2F_1$ is the Hypergeometric function. This solution assumes the infalling boundary conditions at the horizon. It is interesting to note that the same equation, but in a different variable, is obtained in the case of three-dimensional bulk in \cite{Ren:2010ha}. Of course, as the radial variable is different, the two-point function is not the same, as expected from general considerations. By expanding the function (\ref{Hiep}) around $r=\infty$, we can obtain the following analytic expression for the holographic conductivity, 
\begin{align}
    \sigma(\omega)= i \omega\left ( \psi\left(1-\frac{i \omega }{2
   \sqrt{\mu }}\right)+\gamma\right)  +\sqrt{\mu},
\end{align}
where $\psi(x)=\frac{\diff \ln \Gamma(x)}{\diff x}$ is the digamma function, and $\gamma\approx 0.577216$ is Euler's constant. The only relevant scale is the temperature, $T\sim\sqrt{\mu}$. 

\section{Non-minimal coupling conductivity}\label{AppC}

Now we consider the case of non-minimal coupling (\ref{CS_EM_T}). Assuming the same ansatz for the perturbation, $\delta A=e^{-i\omega t}A_x(r)\diff x$,  we again come to a differential equation, but a more complicated one,
\begin{align}\label{nmdiff}\nonumber
    \frac{\diff^2 A_x}{\diff r^2}+ &\frac{3 \lambda^{2} \delta ^2 \mu +3 r^4+r^2 \left(3 \lambda^{2} \delta ^2-\mu \right)}{r \left(r^2-\mu \right) \left(3
   \lambda^{2} \delta ^2+r^2\right)}\frac{\diff A_x}{\diff r}\\&\hspace{25mm}+\frac{\omega^2}{(r^2-\mu)^2}A_x=0.
\end{align}
This equation does not have an analytical solutions, which was to be expected as numerical analysis is usually necessary when dealing with holographic conductivity. In terms of a radial variable $\tz=1/r$ we can  rewrite the equation as 
\begin{equation}
    \frac{\diff^2 A_x}{\diff \tz^2}+\frac{3-\frac{2}{3 \lambda \delta ^2 \tz^2+1}+\frac{2}{\mu  \tz^2-1}}{\tz}\frac{\diff A_x}{\diff \tz} +\frac{\omega^2}{(1-\tz^2\mu)^2}A_x=0.
\end{equation}
To compute the DC conductivity we use the membrane paradigm and take the perturbation $\delta A= (-Et+a(\tz))\diff x$, where $E$ is the boundary electric field while $a(\tz)$ satisfies 
\begin{equation}
    \partial_{\tz}\left(  (1+3\lambda^{2}\delta^2 \tz^2)\left( 1/\tz-\mu \tz \right)\partial_\tz a(\tz)\right)=0,
\end{equation}
and thus quantity $ (1+3\lambda^{2}\delta^2 \tz^2)\left( 1/\tz-\mu \tz \right)\partial_{\tz} a(\tz)$ is conserved in radial evolution. Note that near the boundary ($\tz=0$) we have 
\begin{equation}
    (1+3\lambda^{2}\delta^2 \tz^2)\left( 1/\tz-\mu \tz \right)\partial_{\tz} a(\tz)\rightarrow \frac{1}{\tz}\partial_{\tz}a(\tz)=2A_{x}^{(1)},
\end{equation}
where the identification with $A_{x}^{(1)}$ comes from the structure of the $A_{x}$ assymptotics (\ref{Aasym}).
We impose regularity conditions at the horizon by computing $F_{\mu\nu}F^{\mu\nu}$ to obtain the relation $\partial_{\tz}a(\tz)|_{\tz_h}=\frac{E}{1-\mu\tz^2}|_{\tz_h}$.  This relation enables us to compute the above mentioned conserved quantity at the horizon, yielding 
\begin{equation}
2A_{x}^{(1)}=\left(1+\frac{3\lambda^{2}\delta^2}{\mu}   \right)\sqrt{\mu} E,
\end{equation}
which finally gives us 
\begin{equation}
\sigma_{DC}=\sqrt{\mu}+ \frac{3\lambda^{2}\delta^2}{\sqrt{\mu}}.
\end{equation}

\end{document}